\documentclass[preprint,12pt]{elsarticle}
\usepackage{multicol}
\usepackage{changepage}
\usepackage{lineno,hyperref}
\usepackage{adjustbox}
\usepackage{subcaption}
\usepackage{graphicx}
\usepackage{algcompatible}
\usepackage{amsmath}
\usepackage{graphicx,amsmath}
\usepackage[ruled, lined, longend, linesnumbered]{algorithm2e}
\usepackage{xcolor}
\usepackage{comment}
\usepackage[utf8]{inputenc}
\usepackage[english]{babel}
\usepackage{amssymb}
\usepackage{graphicx}	
\usepackage{algpseudocode}
\usepackage{multirow}
\usepackage{booktabs}
\usepackage{verbatim}

\usepackage{lineno}

\usepackage{hyperref}
\hypersetup{
     colorlinks = true,
     linkcolor = black,
     anchorcolor = black,
     citecolor = black,
     filecolor = black,
     urlcolor = black
     }

\algrenewcommand\alglinenumber[1]{\tiny #1:}
\begin{document}

\begin{frontmatter}

\title{Anomaly-based Framework for Detecting Power Overloading Cyberattacks in Smart Grid AMI}


\author[UBMA]{Abdelaziz Amara Korba}
\author[l3i]{Nouredine Tamani}
\author[l3i]{Yacine Ghamri-Doudane}
\author[ESTI,labged]{Nour El Islem karabadji}

\address[UBMA]{Networks and Systems Laboratory (LRS), Badji Mokhtar-Annaba University, Annaba, Algeria}
\address [ESTI] {Higher School of industrial technologies, Annaba, P.O. Box 218, 23000, Algeria.}
\address[labged]{Electronic Document Management Laboratory (LabGED), Badji Mokhtar-Annaba University, Annaba, Algeria}
\address[l3i]{L3i, Univ. of la Rochelle, La Rochelle, France}

\begin{abstract}
The Advanced Metering Infrastructure (AMI) is one of the key components of the smart grid. It provides interactive services for managing billing and electricity consumption, but it also introduces new vectors for cyberattacks. Although, the devastating and severe impact of power overloading cyberattacks on smart grid AMI, few researches in the literature have addressed them.
In the present paper, we propose a two-level anomaly detection framework based on regression decision trees. The introduced detection approach leverages the regularity and predictability of energy consumption to build reference consumption patterns for the whole neighborhood and each household within it. Using a reference consumption pattern enables detecting power overloading cyberattacks regardless of the attacker’s strategy as they cause a drastic change in the consumption pattern. The continuous two-level monitoring of energy consumption load allows efficient and early detection of cyberattacks. We carried out an extensive experiment on a real-world publicly available energy consumption dataset of 500 customers in Ireland. We extracted, from the raw data, the relevant attributes for training the energy consumption patterns. The evaluation shows that our approach achieves a high detection rate, a low false alarm rate, and superior performances compared to existing solutions.

\end{abstract}


\end{frontmatter}


\section{Introduction}
Information and communication technologies played a crucial role in the growth and performance of the smart grids. The advanced metering infrastructure provides a two-way communication network between smart meters and utility systems, offering interactive services for managing billing and electricity consumption. However, interconnecting the smart grid distributed elements, also introduces new vectors for cyberattacks. The first successful cyberattack on power grid is recorded in December 2015, it struck the Ukraine power grid causing power outages putting more than 100 cities in the dark. The hackers exploited vulnerable points in the infrastructure using a piece of malware known as Black Energy. Several other cyberattacks have followed showing how a hacker with a piece of malware can take control of a power plant’s circuit breaker and damage generators.

Power overloading is one of the most severe cyberattacks, it aims at increasing the energy load to disrupt the load balance on the local power grid, cause a blackout, and damage the grid infrastructure. An attacker with low cost equipment could exploit security vulnerabilities within some points in the smart grid communication infrastructure, particularly within smart meter. The exploit may grant the attacker the command and control of thousands of smart meters that he can subsequently use to dramatically increase the demand of electricity, and to disrupt the load balance on the local power grid. The attacker can also compromise the communication infrastructure
or hack the substation, and then send fake pricing information to the local community. By exploiting the vulnerability of load control systems,
an attacker can modify the consumption profile of the customers and the whole neighborhood (more details are provided in section \ref{sec:power}). Although, the severe impact of power overloading cyberattacks, few works \cite{Jokar2013, liu2014,liu2016,liu2016hi} in the literature have addressed them. 

Traditional anomaly detection systems based on network features did not consider attack scenarios and inherent characteristics of the  smart grid AMI. Current smart grid AMI anomaly detection systems consider fault detection \cite{doi2}, and address mainly two types of cyberattacks: electricity theft and pricing cyberattacks \cite{Giraldo,For, 8172031}. The goal of energy theft is to pay less than the actual price for the consumed energy, in this case the attacker can physically tamper the smart meter, or compromise the communication infrastructure \cite{Zhou2015}. Two common pricing cyberattacks on smart home systems, which manipulate guideline pricing have been studied in the literature \cite{liu2014,liu2016,liu2016hi}. In the first one, named cyberattack for bill reduction, the attacker attempts to fake the guideline pricing curve such that it can reduce the cost of his own bill at the cost of bill increase of other customers. The goal of the second cyberattack is to create a peak energy load in the local community.

Most of the existing anomaly detection systems for smart grid AMI in the literature have been proposed for energy theft detection \cite{Jokar2016, theft1, theft2,theft3, zane, chin, YIP2017} and pricing cyberattacks \cite{liu2014,liu2016,liu2016hi}. However, few researches have addressed overloading cyberattacks targeting grid blackout. Although some works \cite{Jokar2013, liu2014,liu2016,liu2016hi} addressed grid overloading cyberattacks, they only considered short term load increase resulting from pricing manipulation. In this case, the attacker’s goal is to make profit not to shut down the power grid. Anomaly detection systems which monitor the guideline pricing curve are only effective if the attacker overloads the grid by manipulating the guideline pricing curve. Otherwise, the grid overloading cyberattacks would not be detected if the attacker changes its strategy.

Some existing anomaly detection systems proposed in the literature deal with energy theft and grid overloading in the same way. Although energy theft and grid overloading both correspond to an abnormality in the consumption pattern, the two attacks have their proper subtleties and differ in several points such as: attacker’s operating mode, detection delay, and impact on the AMI. Unlike energy theft, grid overloading cyberattack causes an immediate damage, therefore the detection delay is critical in this case. The impact of grid overloading cyberattacks is beyond the smart home, therefore monitoring the energy load at a neighborhood level is needed to avoid cascading failure.

Most of the existing anomaly detection systems in the literature used classification algorithms. One issue with classification-based anomaly detection is the unavailability of malicious samples. Using synthetic malicious samples can solve this issue, however the classifier would not detect unseen attacks that significantly deviate from the synthetic malicious samples used to train the system.

In this paper, we tackle the issue of grid overloading cyberattacks against the smart grid AMI. We propose a consumption pattern-based anomaly detection framework (CPADF) to detect and prevent grid overloading cyberattacks. CPADF applies features engineering and regression-based learning algorithms on historical consumption data to generate normal consumption patterns for the whole neighborhood and for each customer within it. The obtained trained models are then harnessed in a decision making process, we developed, to detect anomalies in consumption patterns. To do so, CPADF monitors continuously electricity consumption at both home and neighborhood level and aggregates anomaly alerts received from customers. An abnormal consumption raise is then detected at a given level if the observed consumption does not match up with its corresponding normal consumption pattern.

We carried out experiments on a real-world publicly available energy consumption dataset of 500 customers in Ireland. We proceeded with data cleaning and feature extraction on the raw data. Initially, the dataset provides 3 attributes only (smart meter identifier, timestamp, and energy consumption) from which we extracted some relevant attributes for training the energy consumption patterns, such as day time, day type, month and season, and we generated labelled datasets for both home and neighborhood levels for the model training and testing\footnote{The generated labelled datasets are made available for free upon request.}. 
The evaluation shows that our approach achieves a high detection rate, a low false alarm rate, and superior performances compared to existing solutions with an optimal training time and memory requirement.

Furthermore, CPADF outperforms existing approach in terms of exploring and detecting sophisticated scenarios of power overloading cyberattacks against smart grid AMI. Indeed, the consumption pattern-based anomaly detection makes CPADF able to detect grid overloading cyberattacks regardless of the attacker’s strategy. Whereas, most of the existing solutions are attacker's strategy oriented, which may fail in detection of cyberattacks if the attacker changes its strategy.


	
	

	

The remainder of this paper is organized as follows. In Section \ref{sec:relWork}, we summarize the state of the art in the field of anomaly detection systems developed to protect smart grid systems. In Section \ref{sec:ami}, we present the AMI network architecture. In Section \ref{sec:power}, different types of power overloading cyberattacks are studied. Section \ref{sec:cpadf} describes the CPADF framework from data collection to anomaly detection process. We evaluate the performance of CPADF in Section \ref{sec:eval}. Section \ref{sec:conc} concludes the paper and draws some lines for future work. 

\section{Related work} \label{sec:relWork}

Most of the existing works in the literature are related to fraud detection, such as electricity theft and pricing cyberattacks. In \cite{liu2014} and \cite{liu2016} the authors considered two smart home pricing cyberattacks: cyberattack for bill reduction; and cyberattack for forming a peak energy load. In the first cyberattack, the hacker attempts to fake the guideline pricing curve such that it can reduce the cost of his own bill at the cost of bill increase of other customers. The goal of the second cyberattack is to create a peak energy usage by faking the guideline pricing curve. A countermeasure technique which uses support vector regression and impact difference for detecting pricing manipulation has been proposed in \cite{liu2014}. The proposed system leverages the interdependence between the electricity pricing and the energy load in the power system. It detects the peak energy load by monitoring changes in the guideline pricing curve. To improve the detection system accuracy, the authors proposed in \cite{liu2016} a partially observable Markov decision process for modeling the long term impact of pricing cyberattacks. In \cite{liu2016hi}, the authors introduced a new type of pricing cyberattack, which creates a sharp increase or decrease of the energy load, resulting in a dramatic drop of generation frequency. To tackle the scalability limitation of the system proposed in \cite{liu2016} and address the new pricing cyberattack, Liu et al.\cite{liu2016hi} proposed a new hierarchical framework, which models the attacking state of each smart meter in a distributed fashion. The proposed framework employs a global policy optimization algorithm to take a centralized decision on checking and repairing the compromised smart meters. In
\cite{liu2014,liu2016,liu2016hi}, the attacker’s objective is to make profit not to cause a blackout by overloading the grid. Although \cite{liu2014,liu2016,liu2016hi} address grid overloading cyberattack,  they only consider short term load increase resulting from pricing manipulation. On the other hand, in this paper we consider different types of long term grid overloading cyberattacks. The proposed anomaly detection systems in \cite{liu2014,liu2016,liu2016hi} are only effective if the attacker creates a peak energy load by manipulating the guideline pricing curve. In this paper, we focus on detecting grid overloading cyberattacks based on consumption pattern changes, regardless of the attacker’s strategy.

Jokar et al. \cite{Jokar2013} addressed grid overloading as well as energy theft. They considered the scenario, where the attacker increases the energy load by manipulating prices or compromising the direct load control system. The authors \cite{Jokar2013} proposed two anomaly detection algorithms based on the predictability of consumption patterns of customers. In \cite{Jokar2016} Jokar et al.  extended and adapted their proposition to detect only energy theft attack. They used transformer meters and anomaly detectors, as well as appropriate classification and clustering techniques, to improve the performance and the robustness of the algorithm against nonmalicious changes in consumption pattern. Classification-based methods need malicious samples to train the classifier, which might not be available, since malicious behavior might never or seldom occur for a given customer. Using synthetic malicious samples can solve the problem. However the classifier would not detect attacks that deviate significantly from the synthetic malicious samples used to train the system. In this paper, we use regression decision trees to predict the consumption profiles during a particular time slot, and then we compare the expected profile with the actual one. Our approach is capable of detecting different attack types, because it does not build the classifier using a particular type of synthetic malicious samples.   

Ford et al. \cite{ford} and Cody et al. \cite{cody} also addressed grid overloading as well as energy theft. They used artificial neural networks and decision tree respectively to model the normal profile of customer’s energy consumption. Real historical data from the Irish smart energy trial \cite{irish} were used to generate the regression models and predict future energy consumption. Then, the anomaly detection systems compare the predicted value with the actual consumption to detect malicious behaviors. Although the proposed approaches \cite{ford} \cite{cody} overcome the limitations of classification-based methods, only one type of grid overloading cyberattack has been considered. The proposed systems do not monitor the consumption pattern at the neighborhood level. Monitoring pattern change at the neighborhood level improves the detection accuracy and reduces the detection delay, since the load increase is more noticeable at the neighborhood level than for a single customer or group of customers, particularly at the beginning of the attack. Important factors such as memory requirement and processing time have not been considered. Using more attributes to model the consumption pattern, CPADF provides a better prediction with lower error rates. CPADF shows good performance in terms of memory requirement and processing time. Our tests show that CPADF outperforms the anomaly detection systems proposed in \cite{ford} \cite{cody}. 

Faisal et al. \cite{stream} proposed a new intrusion detection system (IDS) architecture for the whole AMI system at the levels of smart meter, data concentrator, and headend. A feasibility analysis of the application of several data stream mining algorithms has been conducted to select the best algorithm for each AMI component. In \cite{dids}, Zhang et al. proposed a distributed intrusion detection system for smart grids (SGDIDS) with a hierarchical three layer structure. The proposed IDS analyzes communication traffic using classification algorithms such as support vector machine (SVM) and artificial immune system (AIS). The proposed systems in \cite{dids} and \cite{stream} have been validated on the widely used public KDD Cup 1999 dataset \cite{kdd}. However, this dataset was designed for intrusion detection in computer networks, the considered attacks are based on communication scenarios. The dataset did not consider characteristics inherent to the smart grid infrastructure and attack scenarios against AMI transactions. Furthermore, the KDD dataset \cite{kdd} has a huge number of redundant records and biased distribution of attacks.

In \cite{preventive} an optimal strategy of on-site investigation and monitoring verification for potential anomalies and malware is proposed. Using the decision process framework of Markovian, and based on the observation from the deployed anomaly detectors, the proposed framework determines the best inspection strategies. Alcarez et al. \cite{doi3} examined key security aspects of the Open Charge Point Protocol (OCPP) for communication between electric vehicle, charging points and central management system. The paper shows how a hacker can exploit OCPP vulnerabilities to carry out attacks to burden  resource reservation related to electric vehicle, steal energy, or overload the grid. For instance, an attacker might inject forged OCPP transaction to destabilize network or to affect its functioning. In \cite{doi1} the authors analyzed a set of existing anomaly detection approaches which use machine learning, knowledge and statistical detection-based techniques, and information and spectral theory. The authors investigated the functionalities of the detection approaches for context-awareness in smart grid environments. The paper provides a guideline regarding the choice of the most suitable schemes and detection modes. The suitability is examined based on the restrictions of the context and functional characteristics of the technologies and communication systems. In section \ref{sec:disc}, we show the suitability of CPADF to  the  smart  grid  context according to the set of requirements specified in  \cite{doi1}.

\section {AMI network architecture} \label{sec:ami} 
\begin{figure}	
\begin{center}
\includegraphics[scale=0.35]{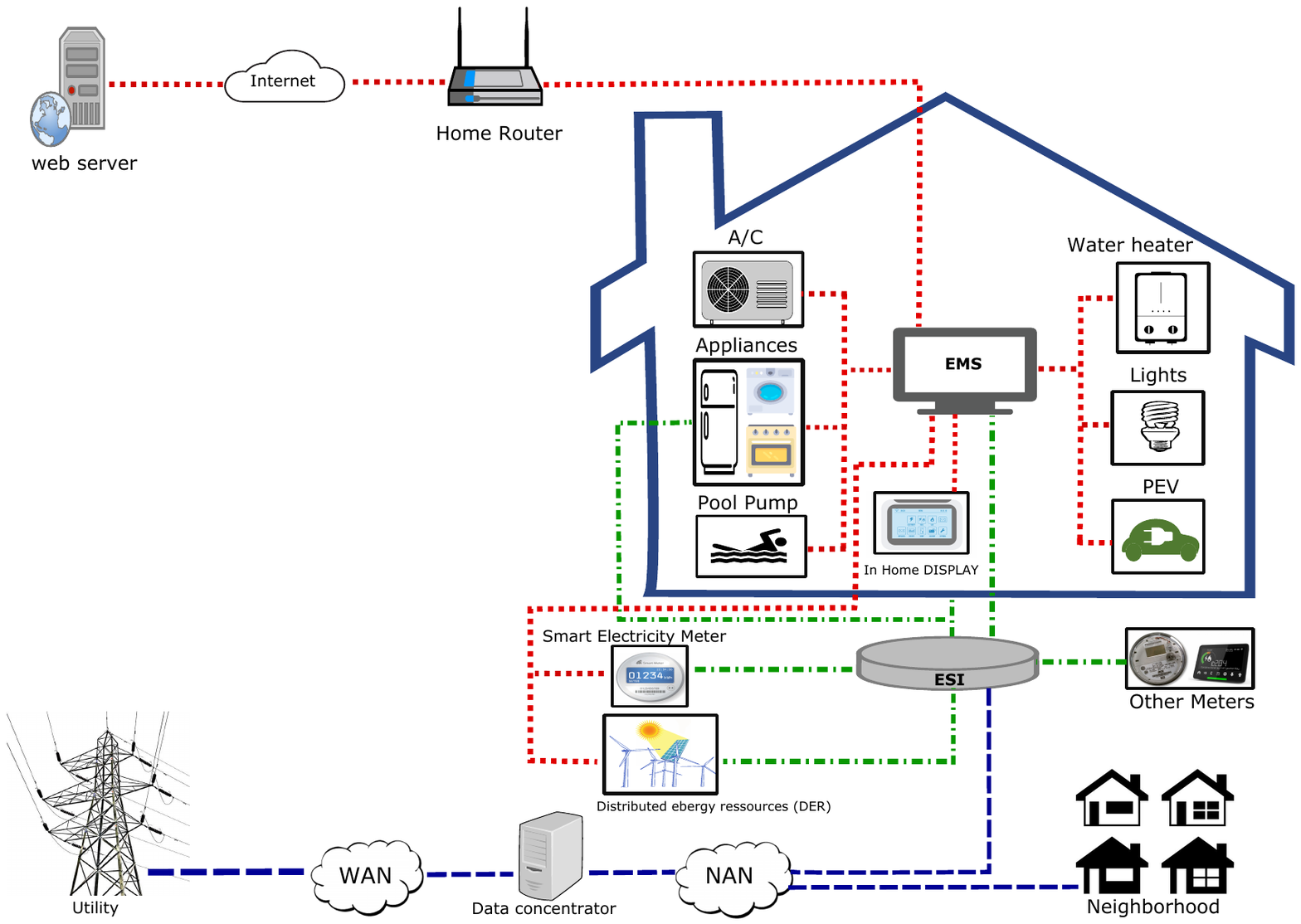}
\end{center}
\caption{AMI network architecture}
\label{smarthome}	
\end{figure}

The smart home (SH) constitutes an integral part of the smart grid AMI, it leverages sensors and networking technologies to be in continuous interaction with its internal and external environments. The Energy Services Interface (ESI) represents the interface connecting the SH to the smart grid. Although there is a logical separation between the smart meter and ESI, their functionalities are generally integrated into one physical device (generally the smart meter) for cost effectiveness. The ESI has diverse functionalities, such as remote control of devices, transmission of consumption data to the utility, supervising of Distributed Energy Resources such as wind turbines, the management of demand response programs, Plug in Electric / Plugin Electric Hybrid Vehicles (PEV/PHEV) charging etc. The Energy Management System (EMS) represents the entity responsible for managing diverse appliances and systems within the SH. It enables the SH to adjust its energy consumption to suit the grid's capacities. The EMS enables the management of high consuming appliances such as air conditioning system, and offers the remote configuration of the smart home devices \cite{komn}. Figure \ref{smarthome} shows the different entities of the AMI network architecture.  
The connections to the ESI are represented by the green dot-dashed lines, whereas the red-dotted lines represent the connections to the EMS. 
The communication between the smart home and the AMI infrastructure is represented by the blue dashed line. The EMS and ESI are in constant two-way communication to manage the internal environment in coherence with the external environment requirements and capabilities \cite{komn}.
The Home Area Network (HAN) interconnects appliances with ESI/smart meters and EMS. The Neighborhood Area Network (NAN) represents the network interconnecting the smart meters with the data concentrator. The Wide Area Networks (WAN) interconnects multiple NANs to the Utility headend.

\section{Power overloading cyberattacks}\label{sec:power}
In this section we present three types of power overloading cyberattacks which exploit the vulnerability of load control systems (such as smart home scheduling systems), and the vulnerability of OCPP protocol. The goal of load control systems is balancing supply and demand to ensure a reliable grid operation. Indirect load control (ILC) mechanisms use dynamic pricing to incite customers to adapt their consumption profiles to suit the grid capabilities. There are two dynamic pricing models, which are usually used together. The first model called real time pricing, where the price is set based on the energy consumption in the local community. The second one is the guideline pricing, where the utility predicts the future load, sets a predictive pricing curve, and uses it for guiding the customers on energy scheduling. The Direct Load Control mechanisms (DLC) allow the utility to directly control the customers’ loads by sending control signals such as turn on/off, through AMI. The OCPP is an application protocol for communication between electric vehicle and charging point and a central management system. One advantage of the introduction of electric vehicles into smart grids, is their bidirectional charging which allows local and global smoothing of imbalances and load peaks. Alcaraz et al. \cite{doi3} studied attacks that misuse the OCPP protocol to destabilize power networks and interfere with resource reservation initiated with the electric vehicle. Although the paper provides divers threat scenarios related to the logical functionality of the OCPP at different stages, in this paper, we consider power overloading scenario at transactions and control stage.

\begin{figure}	
	\begin{center}
		\includegraphics[scale=0.35]{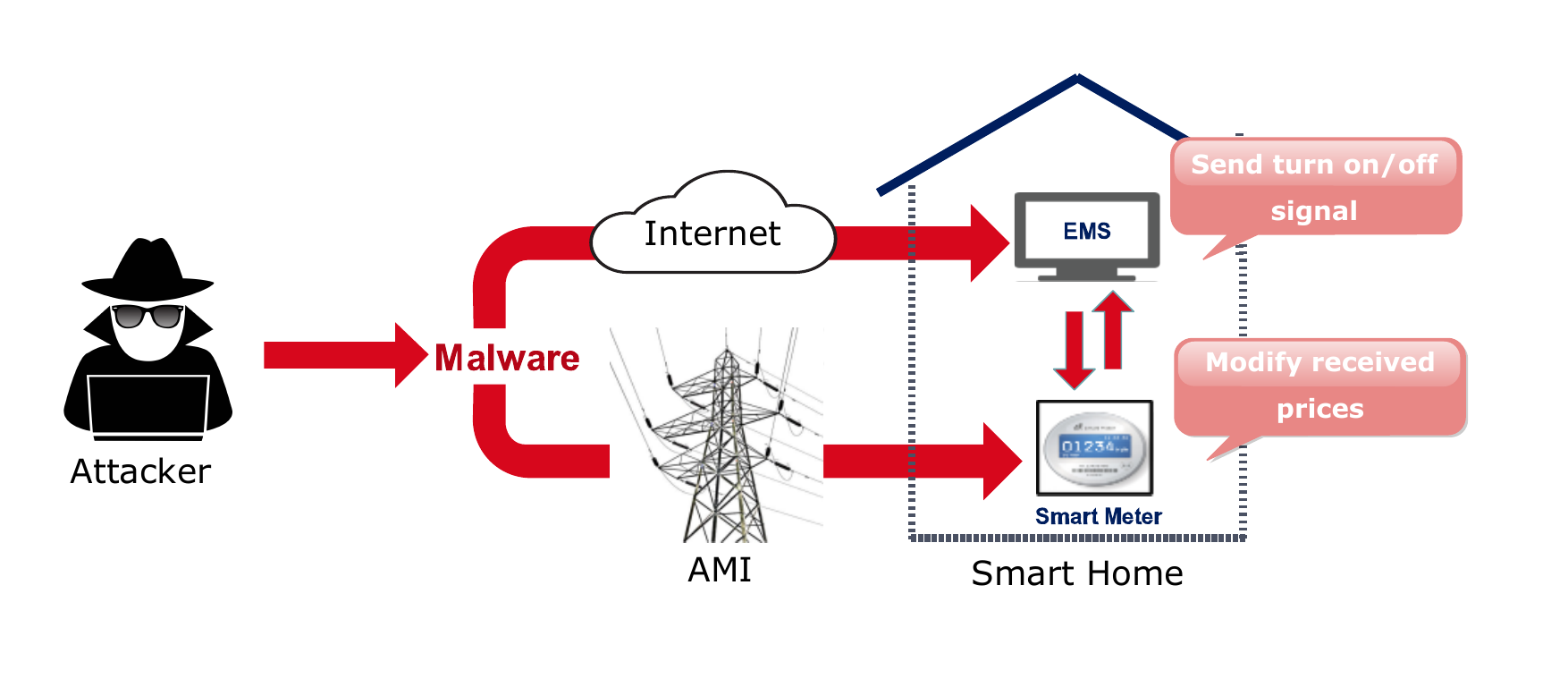}
	\end{center}
	\caption{Power overloading cyberattacks}
	\label{treat}	
\end{figure}

\subsection{Cyberattacks against ILC}
By manipulating the pricing curve, an attacker can modify the consumption profile of the customers and the whole neighborhood consumption profile. The attacker can either compromise the communication infrastructure or hack the substation, and then send fake pricing information to the local community. The attacker can also use a malware to compromise the smart meter and then modify the received pricing information (see figure \ref{treat}). He can then scale that up as much as he may take control of thousands of smart homes, depending on the propagation of the malware \cite{preventive}. In this paper we consider the two following cyberattacks against ILC mechanisms, called pricing cyberattacks \cite{liu2014} as follows.  

\subsubsection{Cyberattack for bill reduction}
The attacker manipulates the guideline pricing curve, in such a way that the electricity price is high during a particular time slot. This will dissuade the other customers to schedule energy consumption during this time slot. Thus, this reduces the local community energy load during this time slot, resulting in the decrease of the real time electricity price there. Afterward, the attacker could schedule the energy consumption during this time slot, and makes profit through reducing his own bill at the cost of bill increase of other customers \cite{liu2014}. 

\subsubsection{Cyberattack for forming a peak energy load}
The attacker first identifies peak consumption energy hours, and then he manipulates the guideline pricing curve such that it is very low during peak energy consumption hours. Therefore, the customers will schedule their large controllable high consumption appliances during peak usage energy hours. This will form a peak in energy consumption leading to significant disturbance in the power system. Also increasing energy load fluctuation could significantly impacts the power system dynamics and changes the generation frequency dramatically. The attacker could increase the energy load fluctuation by manipulating the guideline prices such that it is very high during a time slot then it is very low during the next one, the shorter the time slot is, the higher load fluctuation would be \cite{liu2014} \cite{liu2016hi}.

\subsection{Cyberattacks against DLC} 
The attacker compromises the EMS to send fake “turn on/off” signal ordering a large number of appliances within the premises to get switched on \cite{Jokar2013}. For instance, the attacker can create a surge by turning air conditioners on during peak usage energy periods such as extreme cold/heat or during peak usage hours of the day. Also in this case, the attacker can increase the energy load fluctuation by repeatedly sending turn on/off signals to a large number of appliances, particularly the high consumption ones, such as air conditioning. This will create disturbances and imbalances in the grid that could stumble breakers beyond the targeted neighborhood and cause a large area blackout.
Table \ref{sum} summaries the characteristics of power overloading cyberattacks against load control mechanisms, and shows the anomalous consumption pattern changes.

\subsection{Cyberattack against OCPP}
It has been shown in \cite{doi3} that an attacker may damage the energy safety if the communication channels are intercepted, and the security credentials of an OCPP user/object is known. A hacker might carry out several attacks such as: denial of power resources and services, energy theft, and power overload. As mentioned previously, in this paper, we are interested in power overloading scenario. In smart grids, the majority of charging points are configured to provide bidirectional interfaces for power charging/discharging, so that batteries discharge during peak periods and charge during off-peak times.  The central management system defines the charging profiles which specify the amount of power that can be supplied per time interval to one or multiple points of charge with their charging schedules. To increase power demand at peak periods, the attacker alters the charging profiles, in such a way that the intensity of Wh has to be greater at peak hours or equal to the power consumption in off-peak periods. The fake charging profiles are then used, so that multiple compromised points of charge inject energy into electric vehicle during peak periods.

\begin{table}[]
\centering
	\caption{ Characteristics of power overloading cyberattacks}
	\label{sum}
\begin{adjustbox}{width=1\textwidth}
\begin{tabular}{@{}|l|l|l|l|@{}}
\toprule
\textbf{Cyberattacks} & \textbf{\begin{tabular}[c]{@{}l@{}}Load control \\ mechanisms\end{tabular}} & \textbf{Time slots} & \textbf{Usage patterns} \\ [10pt] \midrule
Bill reduction & ILC & Random & \begin{tabular}[c]{@{}l@{}}Decrease of the whole neighborhood consumption\\ Increase of N compromised smart home consumptions\end{tabular} \\ [10pt] \midrule
Forming peak energy load & ILC/DLC & Peak hours & Increase of the whole neighborhood consumption \\ [10pt] \midrule
Increasing load fluctuation & ILC/DLC & Random and short & Succession of energy consumption increase and decrease \\ [10pt] \bottomrule
\end{tabular}
\end{adjustbox}
\end{table}

\section{CPADF Framework}\label{sec:cpadf}
In this section, the CPADF is described. Firstly, data collection and attributes extraction are described. Next, regression algorithms used for consumption pattern modeling are presented. Lastly, the anomaly detection processes at smart home and neighborhood level are described in details. Hereinafter, we use the abbreviations SH to refer to smart home, and NBH to refer to neighborhood.

\subsection{Data collection and attributes extraction}

\begin{figure}	
	\begin{center}
		\includegraphics[scale=0.4]{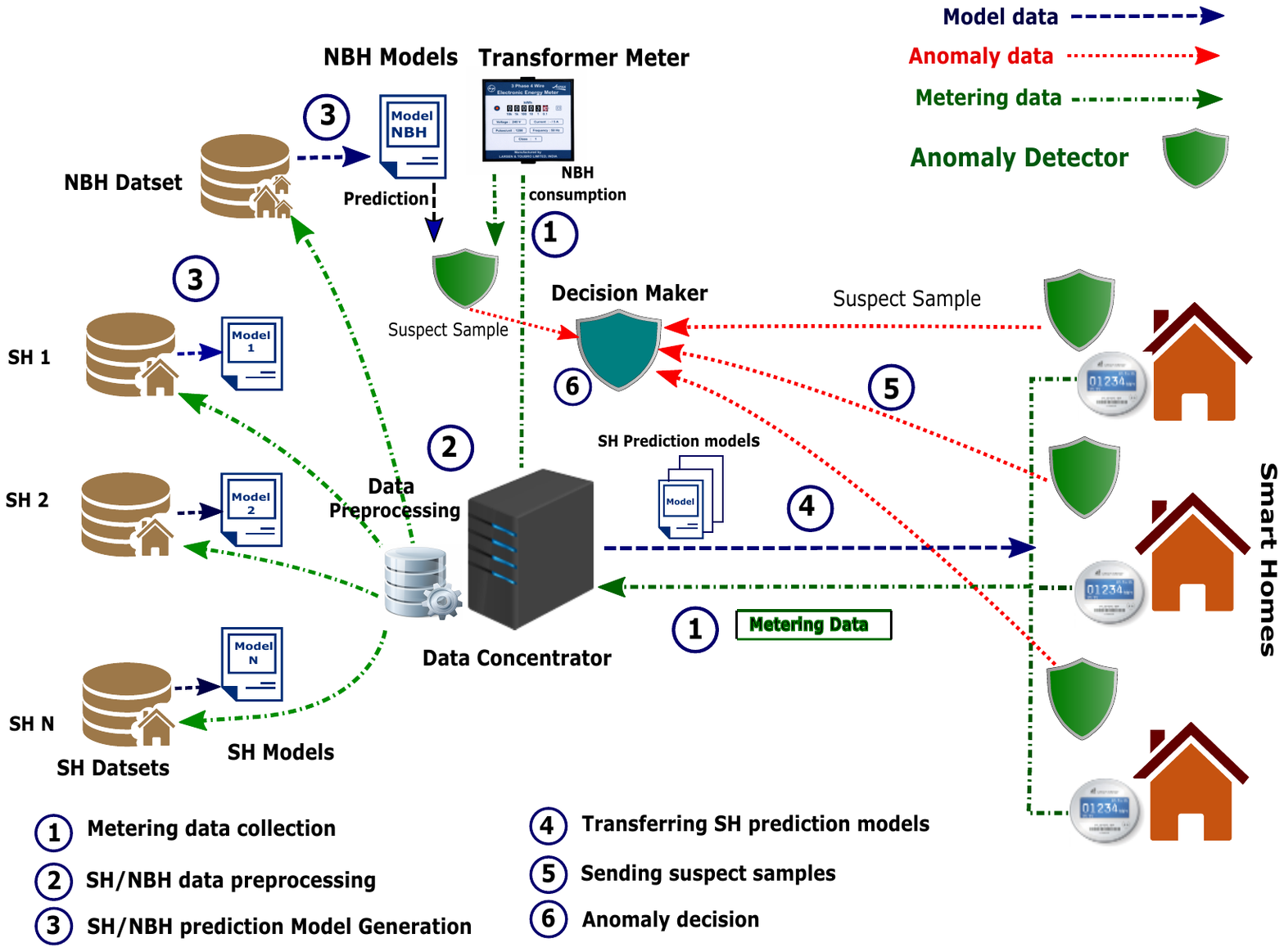}
	\end{center}
	\caption{Data collection, traning and anomaly detection process}
	\label{IDS}	
\end{figure}

Data collection and training process of consumption prediction models for SH and NBH anomaly detectors are illustrated in figure \ref{IDS}. Firstly, metering data are collected from each SH, and from transformer meter. Then, a dataset is generated for each SH, also NBH dataset including the whole neighborhood half hourly consumption is generated. Each data vector within SH and NBH datasets includes the electricity consumption along with a set of time and seasonal related attributes extracted from the raw data (time stamp and consumption). We consider the following attributes: time, day period (day/night), day type (weekday/weekend), month, and season. These attributes are used to allow predicting electricity consumption. For instance, if we consider the attribute day period, most often, electricity consumption tends to decrease during night due to the decline of human activities. The day type attribute allows catching legitimate consumption pattern changes related to the customer activity. For instance, the consumption on the weekend may drop considerably if the customer usually leaves his/her place for some vacations. In contrast, if the customer stays at home, he/she may consume more electricity than usual by spending more time using entertainment devices such as video games, TV, PC, etc. It is important to underline that electricity usage is directly or indirectly affected by external
conditions, particularly by the seasonal conditions as weather and temperature. In winter, the weather is cold and dark, people tend to stay at home, and thus consume more electricity on lighting and heating. While in summer, the weather is sunny and hot, people tend to be out to enjoy sunny weather, so that their electricity consumption decreases. Also, the month attribute needs to be considered, because even within the same season two months could have different consumption pattern, such as September/ December or January/ March. For instance, the consumption pattern of the 1st of October differs from the one of the 21th of December, due to several factors such as: the number of daytime hours and temperature. The NBH prediction model is trained using periodic NBH global consumption calculated and sent by the transformer meter. The SH/NBH historical electricity consumption data are used to model and predict future electricity consumption.The machine learning algorithms used to model consumption pattern are described in the subsequent section \ref{sec:ml}. The SH prediction models are trained within the data concentrator to overcome resource limitation within the ESI. Abnormal consumption samples flagged as suspect by SH anomaly detector are transferred to the data concentrator.

\subsection{Modeling consumption patterns}\label{sec:ml}
To model SH/ NBH electricity consumption pattern so we can predict consumption at any time of the day, five algorithms of supervised machine learning have been used. These algorithms are selected for their known performance and low prediction error rate. The following gives brief description of the machine learning algorithms used in this paper.
\subsubsection{REPTree}
REPTree (Reduced Error Pruning Tree) is a fast decision tree learner that uses information gain ratio (Formula \eqref{eq:gainRatio}) as splitting criterion, where $D$ is the whole dataset, $m$ is the number of classes, $p_{i}$ is the frequency of class $i$ in the dataset, $K$ is the number of subsets generated by the split \cite{Witten2011}.

\begin{equation}
info(D) = \sum\limits_{i = 1}^m - {{p_i}} {\log _2}({p_i})
\label{eq:gainInfo}
\end{equation}

\begin{equation}
GainRatio(A) = \frac{{ info(D) - \sum\limits_{j = 1}^m {\frac{{\left| {{D_j}} \right|}}{{\left| D \right|}}}  \times info ({D_j})}}{{\sum\limits_{j = 1}^k {\frac{{\left| {{D_j}} \right|}}{{\left| D \right|}}}  \times {\log _2}\left( {\frac{{\left| {{D_j}} \right|}}{{\left| D \right|}}} \right)}}
\label{eq:gainRatio}
\end{equation}

\subsubsection{M5P} M5P combines decision tree and linear regression, it uses Standard deviation (SD) to determine the best attribute for splitting the dataset at each node \cite{Witten2011}. The attribute to be chosen is the one that maximizes the error reduction (Formula \eqref{m5p}).
\begin{equation} 
\Delta error=SD(S)-\sum\limits_{i = 1}^m\left(\frac{|D_{i} |}{|D|} SD(D_{i} )\right)
\label{m5p}
\end{equation}

\subsubsection{Random Forest}
Random forest \cite{Breiman2001} is a combination of unpruned  regression trees, it uses random feature selection in the tree induction process. The forest averages the prediction outputs returned by the individual trees.

\subsubsection{Artificial Neural Network}
An artificial neural network is computational system consisting of interconnected simple elements called neurons, which produce output depending on one or more inputs and an activation function e.g., sigmoid function, hyperbolic, etc. Where $\varphi$ in Formula \eqref{ANN} represents the activation function that determines the output value $o$ according to the values of entries $e$ and their weights $w$ \cite{Witten2011}.

\begin{equation}
o = \varphi (\sum\limits_{i = 0}^{i = P} {{w_i}{e_i})}
\label{ANN}
\end{equation}

\noindent 

\subsubsection{SVM}
SVM is a supervised learning model used for classification and regression problems. For regression, SVM uses $\varepsilon$ the insensitive loss function that penalizes error only if it is greater than $\varepsilon$ \cite{Suganyadevi2014}. Therefore, the $\left| \xi  \right|_\varepsilon$ is represented as: 

\[{\left| \xi  \right|_\varepsilon } = \left\{ \begin{array}{l}
0\,\,\,\,if\,\,\,\left| \xi  \right| \le \varepsilon \\
\left| \xi  \right| - \varepsilon \,\,\,\,otherwise.
\end{array} \right.\]

\noindent Using (non-negative) slack variables ${{\xi _i}}$ and ${{\xi _i}^*}$, the final optimization problem to be solved can be formulated as follows: 

\begin{equation}
\begin{split}
Minimize{\kern 1pt} \,\,\frac{1}{2}{\left\| W \right\|^2} + C\sum\limits_{i = 1} {\left( {{\xi _i} + {\xi _i}^*} \right)}
\label{svm2}
\\
Subjected\,to:\hspace{3.5cm}\\
\begin{array}{l}
{y_i} - f({x_i},w) \le \varepsilon  - {\xi _i}^*\\
f({x_i},w) - {y_i} \le \varepsilon  - {\xi _i}^*\\
{\xi _i},{\xi _i}^* \ge 0,\,i = 1,...,n
\end{array}
\end{split}
\end{equation}

\noindent Where $x_{i}$ is a n-dimensional vector, and $y_{i}$ is the target, w is the weight vector, \textit{C} represents the penalty for the error term. SVM regression finds the linear regression in the high-dimension feature space using $\varepsilon$ while reducing the model complexity by minimizing ${\left\| W \right\|^2}$

\noindent The performance evaluation of these algorithms is discussed in the next section (Tables \ref{hanPrec},\ref{nanPrec}).

\begin{figure}[h!]
	\centering
		\includegraphics[scale=0.36]{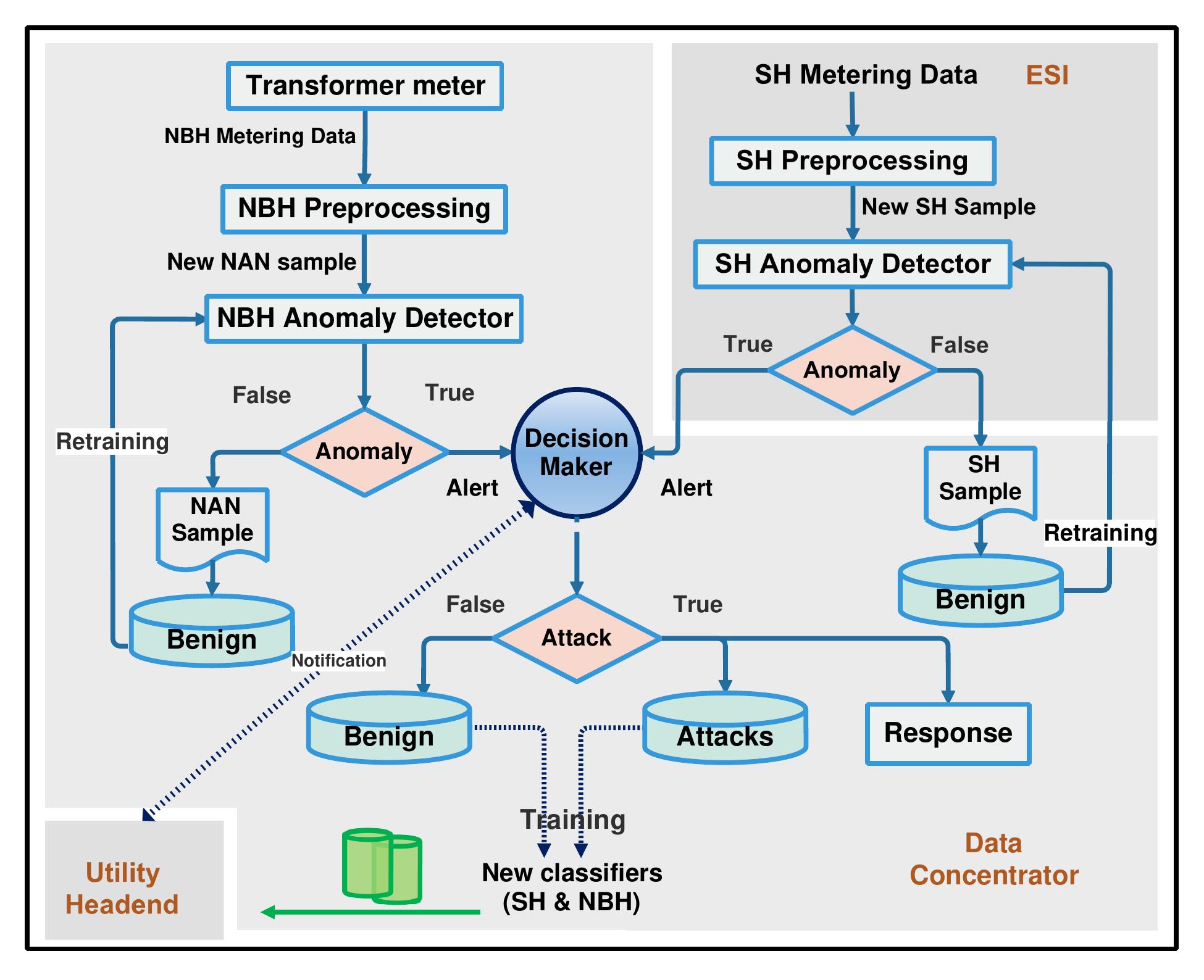}
	\caption{CPADF workflow}
	\label{wf}	
\end{figure}

\subsection{Anomaly detection}
The CPADF two-level monitoring architecture operates within the two main components of the AMI: the ESI, and the data concentrator, as illustrated in Figure \ref{wf}. Periodic metering data from SH are inserted into the SH preprocessing module, which is responsible for attributes extraction and data preprocessing. Since the NBH consumption depends on the consumption of all SHs in the neighborhood, the same set of time related and seasonal attributes are extracted from NBH metering data. The expected consumption calculated by the prediction model, is then compared with the received consumption value.

After preprocessing the SH metering data into the appropriate format consistent with the SH training set, the SH regression model predicts the SH electricity consumption for the given attributes vector. If within $N$ successive time intervals the number of times an anomaly (consumption increase) is detected with a certain threshold ($Nbr_{incr}$), then an anomaly is reported and the suspect sample is sent to the data concentrator. Otherwise, the processed sample is added to the benign dataset which is periodically transferred to data concentrator for periodic retraining. The threshold ($Nbr_{incr}$) specifies the number of tolerable successive abnormal consumption increase. 
This threshold is used to mitigate false alerts caused by occasional legitimate consumption increases. An unusual legitimate consumption increase may be caused by operating one or multiple high energy consumption appliances (washing machine, dishwasher, vacuum cleaning, oven, etc) out of their usual operating time. On average, the length of use of such appliances is between 40 and 120 minutes. For instance, a cycle of washing machine lasts on average between 20 and 60 minutes, most dishwashers cycles are about 2 hours. Thus, the threshold is set to 2 (which corresponds to two successive time intervals of 1 hour). The threshold is set and updated by the system administrator.

\begin{equation}
RMSE=\sqrt{\frac{1}{n} \sum _{j=1}^{n}( \mathop{y}\nolimits_{j} -\mathop{\hat{y}}\nolimits_{j} )^{2} }\label{RMSE}
\end{equation}

 The prediction root mean squared error (RMSE) is used as prediction error (\textit{PE}) for both SH and NBH anomaly detection algorithms. RMSE measures the square root of the average of squared differences between the prediction ($\hat{y}$) and the actual observation ($y$) (see equation \eqref{RMSE}). Since the errors are squared before they are averaged, the RMSE gives a relatively high weight to large errors. By penalizing large errors, the RMSE value increases with the variance of the frequency distribution of error magnitudes. Taking RMSE as prediction error to calculate the threshold, allows to mitigate the detection of legitimate consumption increases as anomalous. The prediction models are retrained when benign datasets are large enough in terms of instances. Periodic retraining allows CPADF to adapt to consumption pattern changes related to nonmalicious factors such as changes of residents or appliances, etc. Pseudocodes of the SH anomaly detection algorithm is provided in Algorithm \ref{alg:sh}.


\begin{algorithm}[h!]
    \caption{SH Anomaly detection algorithm}
    \label{alg:sh}
     \textbf{BEGIN}
     
     \textbf{Input}: \textit{$SH_{meter}$ (SH metering data)} \;
      \textbf{Output}: \textit{anomaly (boolean)}\;
      
      \textbf{Variables}: \textit{t (time interval), $Pred_{SH}$ (SH prediction model), $SH_{Cons}$ (SH observed consumption), counter (number of times an increase is detected),$SH_{PE}$ (prediction error), $Nbr_{incr}$ (threshold of successive consumption increase), $BD_{SH}$ (SH benign dataset)} \; 

\For{$t \in \{1,...,24\}$}{

  	Calculate attributes vector $NS_{SH}$ from $SH_{meter}$}

  	$SH_{PC}=Pred_{SH}$($NS_{SH}$)\;
  
   \eIf{( $SH_{Cons}>SH_{PC}+SH_{PE}$)}{%
    		\eIf{($counter>Nbr_{incr}$) during the last N time intervals)}{
    		        \textit{anomaly}=true \;
    		        \textit{Send alert to decision maker} \;   
    		        Transfer the suspect sample to the data concentrator \;
    		        }{
    		        \textit{counter++}\;
    		        \textit{Add $NS_{SH}$ to $BD_{SH}$} \;
    		        } }{
    		        
    		            \textit{Add $NS_{SH}$ to $BD_{SH}$}\; 
    		        
    		        }
    		         
  	Periodic transfer of $BD_{SH}$ to the data concentrator \;
  	 \textbf{END}
\end{algorithm}

At the SH level, the periodic consumption monitoring is set to 1 hour, because humans typically operate on hour interval, therefore it is difficult to notice pattern change over a smaller time interval. However, consumption pattern changes of a large number of SHs even over a shorter period of time results in drastic consumption pattern change of the whole NBH. Therefore, at the NBH level, the periodic consumption monitoring is set to the smart meter data collection frequency (30 minutes) to provide the minimal detection delay. The NBH total electricity is measured by the transformer meter. After calculating and preprocessing the NBH attributes vector, this latter is given to the NBH regression model to predict consumption. If the received consumption is larger than the sum of the predicted consumption and the prediction error (RMSE), then a Neighborhood Abnormal Consumption Raise (NACR) is detected, the suspect sample is stored, and an alert is sent to the decision maker. Otherwise, the processed sample is added to the benign dataset ($BD_{NBH}$) for the periodic retraining of NBH consumption profile. Pseudocodes of the NBH anomaly detection algorithm is provided in Algorithm \ref{alg:nbh}.

\begin{algorithm}[h!]
    \caption{NBH Anomaly Detection Algorithm}
    \label{alg:nbh}
   \textbf{BEGIN}
   
   \textbf{Input}: \textit{ $NBH_{meter}$ (NBH metering data)} 
   
	\textbf{Output}:  NACR (boolean)
		
	\textbf{Variables}: \textit{t (time interval), $Pred_{NBH}$ (NBH prediction model),$NBH_{PE}$ (prediction error), $NBH_{Cons}$ (NBH observed consumption), $BD_{NBH}$ (NBH benign Dataset)} 
     
  \For{$t \in \{1,...,48\}$}{
  Calculate attributes vector $NS_{NBH}$ from $NBH_{meter}$ \;
  $NBH_{PC}=Pred_{NBH}$($NS_{NBH}$)\;

    \eIf{$(NBH_{Cons}>NBH_{PC}+NBH_{PE})$} {%
    		
    	NACR=true \;
		Send alert to the decision maker \;
		Store the suspect sample \;

    		}{
    		        
    		  NACR=false \;
		      Add $NS_{NBH}$ to $BD_{NBH}$;\\
    		        } }
    Periodic training of $Pred_{NBH}$  \;
\textbf{END}
\end{algorithm}

The decision maker module confirms the anomaly and notifies the operator in two cases: 
1) more than half of SH anomaly detection systems in the NBH report an anomaly; 2) when a neighborhood abnormal consumption raise is reported. The first case corresponds to bill reduction cyberattack, the second one corresponds to forming peak energy load cyberattack. Then, the utility headend checks whether the detected anomaly is caused by a cyberattack or it is related to a temporary pattern change such as special occasions. After decision, the appropriate response is triggered, the attack or normal samples are stored into either attack or benign datasets. Initially, attack datasets are empty
unless external sources are used. Malicious samples classified by the decision maker will be added to the attack datasets. Once the two datasets are
large enough, they will be used to build new classifiers for SH and NBH.
These classifiers will constitute a second detection level and a decision support system for the utility headend. This approach allows for overcoming issues
related to using synthetic malicious samples to train the classifiers. If NACR
was detected, but no anomalies have been reported, it appears that an attack
might be occurring but the SH anomaly detection system cannot recognize
it. In this situation, the SH dataset is analyzed for sign of gradual overloading
cyberattack, in which the attacker gradually increases the consumption data
to mislead the learning machine to consider a malicious pattern as a normal
one. The long-term tendency in daily consumption (historical data) of the smart
home is analyzed. A gradual overloading can be characterized by an ascending
slope in long-term consumption curve.
Pseudocodes of the decision making algorithm is provided in Algorithm \ref{alg:dm}.

\begin{algorithm}[h!]
    \caption{Decision Making Algorithm}
    \label{alg:dm}
   \textbf{BEGIN}
   
   \textbf{Input}: \textit{ NACR (boolean), $Nb_{alert}$ } 
   
	\textbf{Output}: attack (boolean)

	\textbf{Variables}: \textit{$Nb_{SH}$ (number of SH in the neighborhood)}

\For{$t \in \{1,...,48\}$}{ 
    	    	
    	    	   \eIf{$(NACR==true) || (Nb_{alert}>\frac{1}{2}*Nb_{SH}$)}{
    	   attack=true \;
		   Send an alert to the operator \;
    	       	   }
    	{     Add $NS_{NBH}/NS_{SH}$  to $BD_{NBH}/BD_{SH}$ \;

    	}

    	    }
    	    
    	   \eIf{(attack is confirmed)}{
    	   Add $NS_{NBH}/NS_{SH}$ to the attack datasets \;
	       Initiate a response \;
    	       	   }
    	{     Add $NS_{NBH}/NS_{SH}$  to $BD_{NBH}/BD_{SH}$ \;

    	}
    	
    \textbf{END}
\end{algorithm}

\begin{table}[]
	\centering
	\caption{Raw data file format}
	\label{raw}
	\begin{tabular}{c|c|c}
		\hline
		Meter ID & Encoded date/time & Energy Consumption (KWh) \\ \hline
		1392 & 19535 & 0.256 \\ \hline
		1392 & 19538 & 0.265 \\ \hline
		1951 & 19604 & 0.042 \\ \hline
		1951 & 19605 & 0.021 \\ \hline
	\end{tabular}
\end{table}
\section{Experimental results} \label{sec:eval}
We used in our experimentation the smart meter energy consumption data from the Irish Smart Energy Trial \cite{irish}, the dataset was released by SEAI in January 2012. The dataset has been created within Smart Metering Electricity Customer Behaviour Trials (CBTs) which has taken place from 2009 to 2010. The purpose of the trials was to assess the impact on consumer’s electricity consumption in order to inform the cost-benefit analysis for a national rollout. The dataset contains the energy consumption data of over 5000 residential households and businesses \cite{irish}. The dataset is constituted of six data files with millions of entries per file. Each data file contains three columns, the first column indicates the smart meter ID which identifies a particular resident or business. The second column represents timestamps corresponding to the time and date of the meter reading. Digits 1-3 represents the day code (day 1 = 1st January 2009), time code is represented by digits 4-5 (1-48 for each 30 minutes with 1= 00:00:00 – 00:29:59). The third column indicates the energy consumption value in kilowatt-hours (kWh). Table \ref{raw} shows a small sample of the raw data.

\subsection{Datasets generation and preprocessing}
The raw dataset includes the energy consumption data of all customers. To model each customer's consumption pattern separately, the raw consumption data are split by meter ID into a collection of consumption datasets. For each customer dataset, a set of attributes are generated. Each vector in the new dataset includes the following attributes: SH consumption per hour; hour (1, ..., 24); day type (weekday or weekend); month and season. Among four consecutive weeks one week is randomly chosen for the validation set and the other 3 weeks for the training set.  Thus we use 75 \% of the dataset for training and 25 \% for validation. Likewise, from the raw data the NBH dataset is generated. Each vector in NBH dataset consists of the half hourly consumption of the whole NBH (the summation of meter reading of all customers within the neighborhood) and the same attributes used for SH datasets: hour, day type, day period, month and season. Among four consecutive weeks one week is randomly chosen for the validation set and the other 3 weeks for the training set.

\begin{figure}
    \centering
    \includegraphics[scale=0.7]{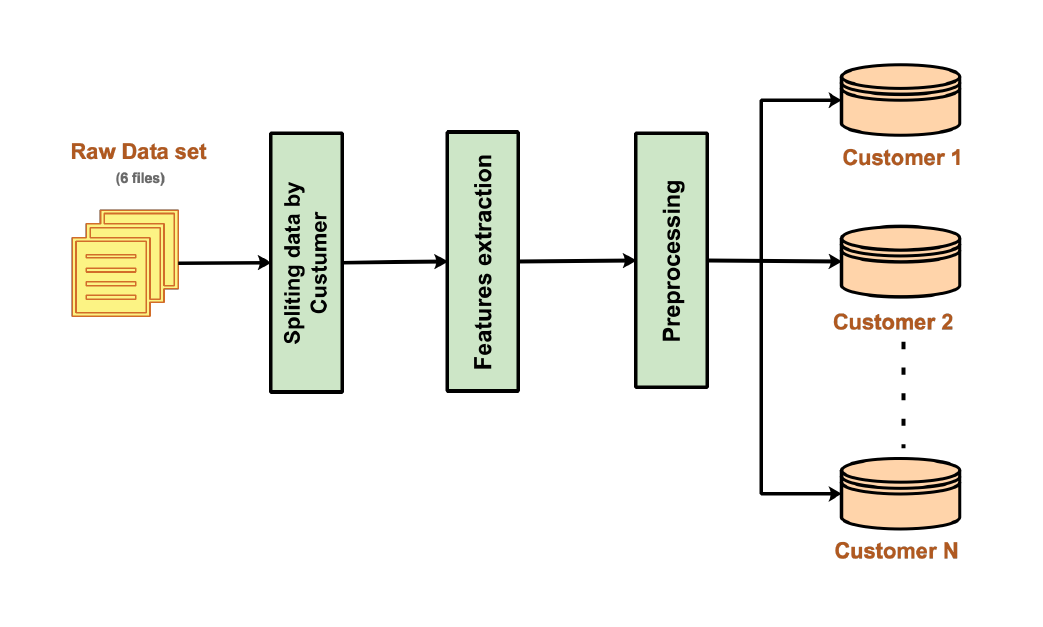}
    \caption{Datasets generation process}
    \label{fig:prepross}
\end{figure}

Data preprocessing includes operations such as cleaning and normalization. The cleaning task consists in identifying missing values and eliminating outliers and extreme values. Outliers and extreme values such as peak energy consumption may correspond to unusual activities such as holidays or special occasions. The mean and standard deviation $\sigma$ for each time interval within each month is calculated. All consumption values that do not lie within three $\sigma$ of the mean are removed from the dataset. Figure \ref{fig:prepross} summarizes the datasets generation process. As indicated previously, different time intervals have been used for SH and NBH datasets because at SH level it is difficult to notice pattern change over a small time interval, humans typically operate on hour interval. However, at NBH level the pattern change can be noticed.

\subsection{Energy consumption prediction}
We have used Weka \cite{weka} in our experiment, it is a collection of open source machine learning algorithms for data mining tasks. The performance of each of the five algorithms discussed in the previous section is measured in terms of the following metrics: 
\begin{enumerate}
	
\item Mean Absolute Error (MAE): measures the average absolute differences between the predicted value and the actual value in the validation dataset (see Equation \eqref{MAE}).

\item Root Mean Squared Error (RMSE): measures the square root of the average of squared differences between the predicted value and the actual value in the validation dataset (explained previously) \eqref{RMSE}).

\item Running time (in seconds): the time taken to build the model
\item Model size (KB): the size of the prediction model in kilobytes

\end{enumerate} 

\begin{equation}
MAE=\frac{1}{n} \sum _{j=1}^{n}\left|\mathop{y}\nolimits_{j} \right.  -\left. \mathop{\hat{y}}\nolimits_{j} \right|
\label{MAE}
\end{equation}

We can see from the results of the energy consumption prediction of 500 customers in Table \ref{hanPrec}, that M5P algorithm gives the smallest average error rate within a reasonable running time and with low memory requirement. Therefore, M5P constitutes the best algorithm to use for SH energy prediction. The MLP algorithm shows the highest error rates and the longest running time. The random forest presents a huge memory usage in comparison with the other algorithms. Concerning the NBH energy prediction, table \ref{nanPrec} shows that REPTree algorithm provides the best trade-off between error rates, running time and memory requirement. Therefore, we choose REPTree algorithm for NHB energy prediction, and M5P algorithm for SH energy prediction.    
  
\begin{table}[]
	\centering
	\caption{Performances of regression algorithms on SHs data}
	\label{hanPrec}
	\begin{tabular}{@{}lllll@{}}
		\toprule
		Algorithm & MAE & RMSE & Running Time (s) & Model Size (KB) \\ \midrule
		REPTree & 0.395 & 0.534 & 0.123 & 9.979 \\
		M5P & 0.329 & \textbf{0.453} & 0.724 & 13.273 \\
		RandomForest & 0.403 & 0.549 & 3.366 & 3,810.904 \\
		SVM & 0.372 & 0.508 & 6.132 & 110.735 \\
		MLP & 0.426 & 0.562 & 17.263 & 16.286 \\
		\bottomrule
	\end{tabular}
\end{table}

\begin{table}[]
	\centering
	\caption{Performances of regression algorithms on NBH data}
	\label{nanPrec}
	\begin{tabular}{@{}lllll@{}}
		\toprule
		Algorithm & MAE & RMSE & Running Time (s) & Model Size (KB) \\ \midrule
		REPTree & 351,803 & \textbf{478,770} & 5,000 & 113,000 \\
		M5P & 353,312 & 480,546 & 21,000 & 86,000 \\ 
		Random Forest & 350,087 & \textbf{476,013} & 59,000 & 11386,000 \\ 
		SVM & 585,336 & 773,824 & 12832,000 & 2709,000 \\ 
		MLP & 451,850 & 582,765 & 246,000 & 15,000 \\ 
		\bottomrule
	\end{tabular}
\end{table}

\subsection{Overloading cyberattacks detection}
To the best of our knowledge, no real smart grid AMI transaction dataset including overloading cyberattacks data is publicly available. Thus, we simulate the power overloading cyberattacks against ILC/DLC discussed in Section \ref{sec:cpadf} for 500 customers. We implement theses attacks based on datasets of normal samples, for each instance of the dataset we generate four types of malicious samples as follows (refer to table \ref{tab:var} for variable description): 
\begin{enumerate}
	\item Attack of type 1: this attack simulates forming peak energy load, where the attacker attempts to overload the grid during times of high demand when the grid becomes under pressure:
\[\begin{array}{l}
{M_1}(e) = e + {\alpha _t}\\
\mathop \alpha \nolimits_t  = \left\{ \begin{array}{l}
random(0.8,4),Pea{k_{start}} \le t \le Pea{k_{end}}\\
1\;\;otherwise
\end{array} \right.\\
Peak\;hours:\{ 7 - 9\} \{ 19 - 22\} 
\end{array}\]	
Where \textit{e} is the normal consumption value and \textit{M} is the modified consumption value.

	\item Attack of type 2: this attack simulates bill reduction cyberattack, where the attacker manipulates the guideline price with a low price from $tim{e_{start}}$ to $tim{e_{end}}$ to urge customers in the community to schedule energy during this period, and a high price at other time slots during which he can schedule his energy load. Thus, the energy load increases from $tim{e_{start}}$ to $tim{e_{end}}$.
\[\begin{array}{l}
{M_2}(e) = {\beta _t} + e\\
{\beta _t} = \left\{ \begin{array}{l}
(0.8,4),\;tim{e_{start}} \le t \le tim{e_{end}}\\
1\;\;otherwise
\end{array} \right.\\
tim{e_{start}} = random(0,23 - \min OffTime)\\
duration = random(\min OffTime,24)\\
tim{e_{end}} = tim{e_{end}} + duration\\
\min OffTime = 4;
\end{array}\]
		
	\item Attack of type 3: to ensure a higher impact, the attacker may attempt to create a sharp energy increase. This attack simulates a variant of forming peak energy cyberattack. In this case, the amount of energy increase is greater than in the case of attack of type 1.
	
	\item Attack of type 4: this attack simulates increasing load fluctuation cyberattack. The attacker alternates repeatedly between normal behavior and grid overloading to disturb the grid.
	
\end{enumerate}

\begin{figure}[h!]
\centering
  \begin{subfigure}[b]{0.555\textwidth}
    \includegraphics[width=\textwidth]{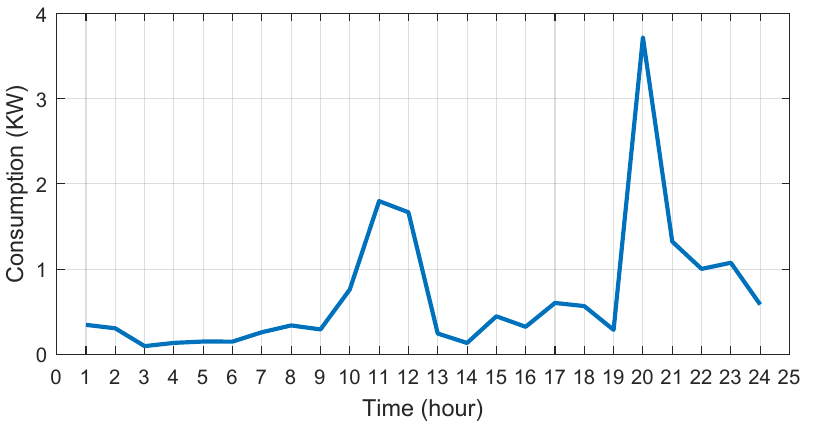}
    \caption{Normal consumption pattern}
    \label{norm}
  \end{subfigure}
  \begin{subfigure}[b]{0.558\textwidth}
    \includegraphics[width=\textwidth]{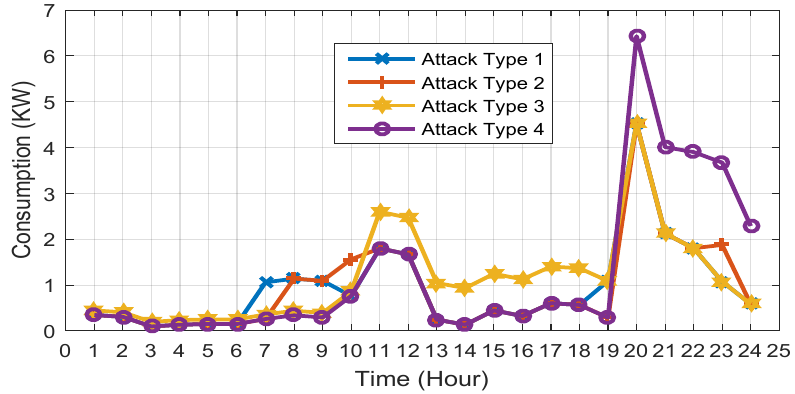}
    \caption{Attack consumption patterns}
    \label{attack}
  \end{subfigure}
  \caption{A sample one-day period of energy consumption}
 \end{figure}
 
\begin{table}[]

\centering
\caption{Variables description}
\begin{adjustbox}{width=0.9\textwidth}
\begin{tabular}{@{}ll@{}}
\toprule
\textbf{Variables}       & \textbf{Descriptions}                                            \\ \midrule
M1                       & Malicious consumption pattern generated through attack of type 1 \\
M2                       & Malicious consumption pattern generated through attack of type 2 \\
$e$                       & Normal consumption                                               \\
$Peak_{start}$          & starting time of peak hours, e.g.  7 am                           \\
$Peak_{end}$            & ending time of peak hours, e.g., 10 pm                           \\
$\alpha_t$ & the amount of electricity increase in attack of type 1           \\
$\beta_t$ & the amount of electricity increase in attack of type 2           \\
$time_{start}$          & attack of type 2 starting time                                   \\
$time_{end}$            & attack of type 2 ending time                                     \\
minOffTime               & attack of type 2 minimal duration, e.g. 4 hours                  \\ \bottomrule
\end{tabular}
\end{adjustbox}
\label{tab:var}
\end{table}

\noindent Figure \ref{norm} shows an example of the energy consumption of a particular customer during a single day. Figure \ref{attack} illustrates the corresponding attack patterns.

\noindent We simulated the same four attack types on the neighborhood dataset, we implemented the attacks based on NBH dataset of normal samples. For each instance of the NBH dataset, we generated four types of malicious samples in the same way as described previously. We adjust the amount of energy consumption increase based on the neighborhood average consumption. The performance of the anomaly detection algorithms is measured in terms of the following metrics: Accuracy (AC); True Positive rate (TPR); False Positive Rate (FPR); True Negative Rate (TNR); and False Negative Rate (FNR). As we can see in Table \ref{SHAD}, the average RMSE on SH attack samples (RMSE-A) deviates considerably from the average RMSE on SH normal samples (RMSE) regardless of the attack type. The deviation is more considerable in the case of attack of types 3 and 4 because the amount of energy increase is more important. The SH anomaly detection algorithm shows high detection performances. It delivers high accuracy and detection rate with low false positive and false negative rates. We observe the best detection rate on attack of type 4, and the lowest false positive rate on attack of type 1. 

\begin{figure}[h!]
\centering
  \begin{subfigure}[b]{0.45\textwidth}
    \includegraphics[width=\textwidth]{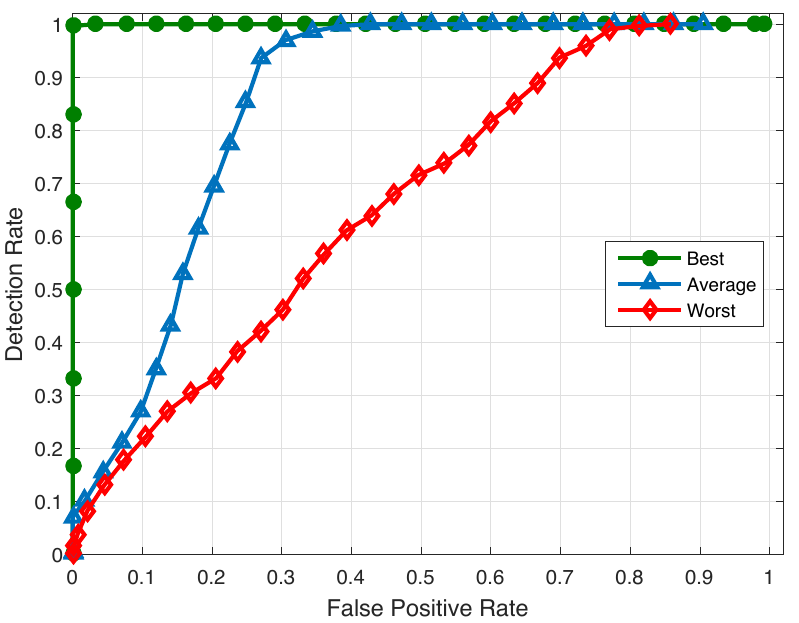}
    \caption{Attack of type 1}
    \label{fig:T1}
  \end{subfigure}
  \begin{subfigure}[b]{0.45\textwidth}
    \includegraphics[width=\textwidth]{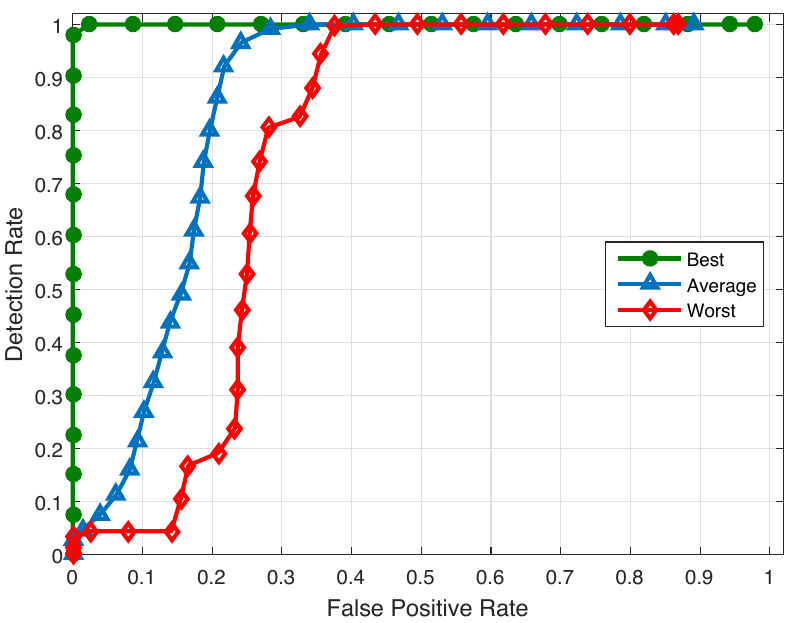}
    \caption{Attack of type 2}
    \label{fig:T2}
  \end{subfigure}
  \begin{subfigure}[b]{0.45\textwidth}
    \includegraphics[width=\textwidth]{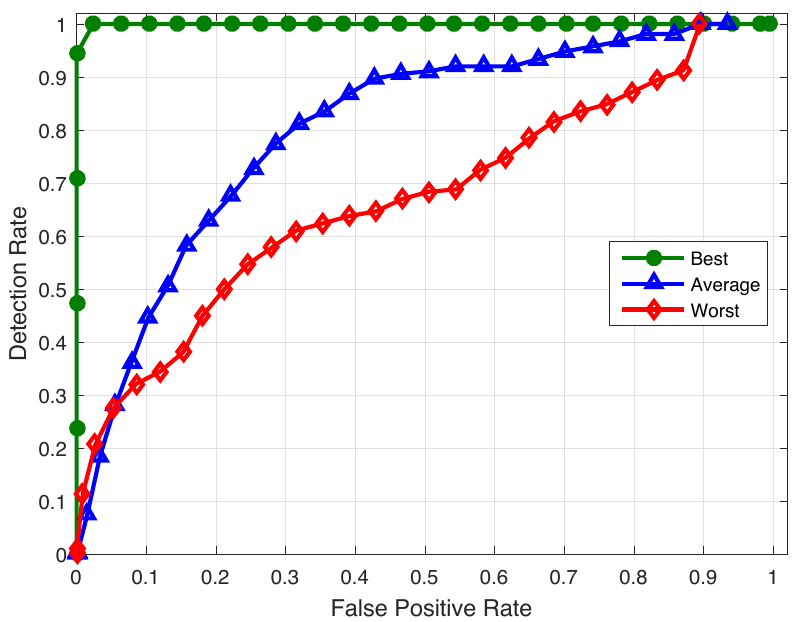}
    \caption{Attack of type 3}
    \label{fig:T3}
  \end{subfigure}
  \begin{subfigure}[b]{0.45\textwidth}
    \includegraphics[width=\textwidth]{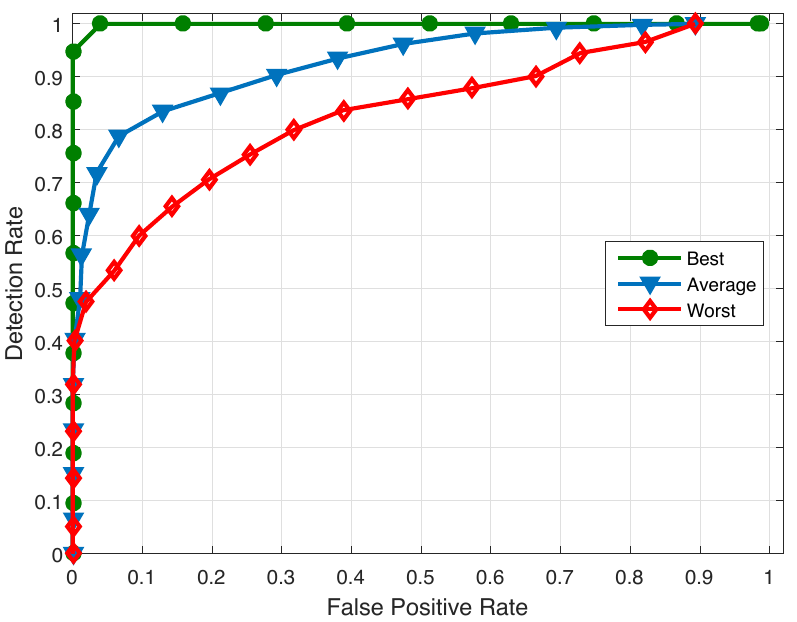}
    \caption{Attack of type 4}
    \label{fig:T4}
  \end{subfigure}
  \begin{subfigure}[b]{0.45\textwidth}
    \includegraphics[width=\textwidth]{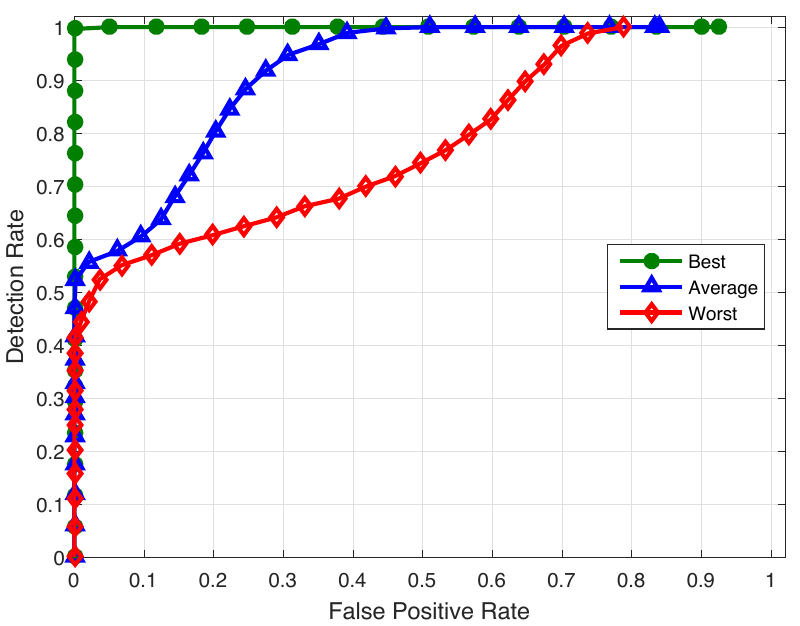}
    \caption{All attacks}
    \label{fig:all}
  \end{subfigure}
  \caption{ROC curves of SH anomaly detection system}
  \label{fig:rocs}
\end{figure}

To highlight the trade-off between TPR and FPR, we relied on the Receiver Operator Characteristic (ROC) curve which
plots the TPR (y-axis) against the FPR (x-axis). Figure \ref{fig:rocs} shows the ROC curves of three customers with best, intermediate, and worst performances of attack of types 1, 2, 3, 4, and all the attack types combined, respectively. As we can notice, the curves are closer to the top-left corner indicating a good performance on detecting the four attack types combined or separated. The ROC curve \ref{fig:T4} confirms that CPADF delivers the best detection performances on attack of type 4. Figure \ref{fig:all} shows the capacity of CPADF to maintain high detection rate with low false positive rate against all attack types combined. A summary of the NBH anomaly detection results is listed in Table \ref{NBHAD}, the NBH detection algorithm shows high detection performance. We observe the best detection rate on attack of type 3, and the lowest false positive rate on attack of type 1.

\begin{table}[h!]
	\centering
	\caption{SHs Anomaly Detection.}
	\label{SHAD}
	\begin{tabular}{@{}llllllll@{}}
		\toprule
		& RMSE & RMSE-A & AC & TPR & FPR & TNR & FNR \\ \midrule
		Type 1 & 0.332 & 0.965 & 0.901 & 0.918 & 0.104 & 0.896 & 0.082 \\
		Type 2 & 0.332 & 0.942 & 0.903 & 0.920 & 0.111 & 0.889 & 0.080 \\
		Type 3 & 0.332 & 2.625 & 0.924 & \textbf{0.989} & 0.120 & 0.880 & 0.011 \\
		Type 4 & 0.332 & 2.617 & 0.947 & \textbf{0.991} & 0.120 & 0.880 & 0.009 \\
		All & 0.332 & 2.673 & 0.907 & 0.965 & 0.161 & 0.839 & 0.035 \\ \bottomrule
	\end{tabular}
\end{table}
\begin{table}[h!]
	\centering
	\caption{NBH Anomaly Detection}
	\label{NBHAD}
	\begin{tabular}{@{}llllllll@{}}
		\toprule
		& RMSE & RMSE-A & AC & TPR & FPR & TNR & FNR \\ \midrule
		Type 1 & 478.770 & 1168.923 & 0.890 & 0.882 & 0.108 & 0.892 & 0.118 \\
		Type 2 & 478.770 & 1130.237 & 0.893 & 0.909 & 0.121 & 0.879 & 0.091 \\
		Type 3 & 478.770 & 2678.249 & 0.899 & \textbf{0.998} & 0.168 & 0.832 & 0.002 \\
		Type 4 & 478.770 & 2687.301 & 0.931 & \textbf{0.993} & 0.168 & 0.832 & 0.007 \\
		All & 478.770 & 2923.112 & 0.901 & 0.958 & 0.166 & 0.834 & 0.042 \\ \bottomrule
	\end{tabular}
\end{table}

\subsection{Discussion and comparison} \label{sec:disc}

\begin{figure}[h!]
	\begin{center}
		\includegraphics[scale=0.7]{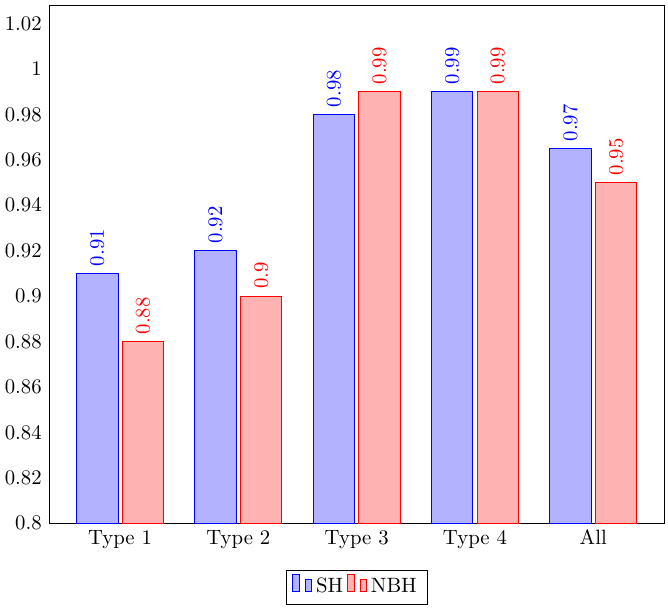}
	\end{center}
	\caption{Comparison of TPR among attack types }
	\label{fig:TPR}	
\end{figure}

The CPADF shows high accuracy on detecting the four attack types combined or separated, at both SH and NBH levels. However, in the context of smart grids, the two classes (attack and normal) are not equally important. It is known that TPR would be the metric to use when there is a high impact associated with false negative (attack classified as normal). It is safer for the system to tolerate false positive (normal consumption change detected as attack) rather than false negative. The impact of false negative would be extremely high if the target system is connected to other systems (cascading failures). Against all attacks combined, the TPR is higher than 96 \%, and the FNR is less than 4 \%. The superior TPR on detecting attack types 3 and 4 shows the effectiveness of CPADF when there are drastic changes in consumption patterns, as illustrated in Figure \ref{fig:TPR}. The highest FNR is noticed on attack of type 1, due the fact that during peak hours, differentiating between legitimate and malicious consumption increase is more challenging. Furthermore, this is in part because the random generation of the amount of energy increase can in some cases return consumption values which are close to normal consumption values. Due the aforementioned facts, a slight drop in TPR at NBH level can be observed in the cases of attack types 1 and 2 (see Figure \ref{fig:TPR}). The results showed the effectiveness and high performances of CPADF on detecting different types of overloading cyberattacks at SH and NBH levels.

According to \cite{doi1} an anomaly-based detection system must fulfil a set of requirements to be suitable to the smart grid context: operational performance ([R1]); reliability and integrity in the control ([R2]); resilience ([R3]); security ([R4]) and privacy ([R5]). CPADF complies with security and resilience requirements ([R3, R4, R5]) thanks to the periodic retraining ensuring incremental learning to update the knowledge of the system with new legitimate consumption patterns. Using RMSE in threshold calculation allows controlling subtle changes, while two-level monitoring (home and neighborhood) of the consumption load enables controlling drastic changes in electricity consumption and load demand. The decision trees low computational complexity and fast learning, along with their comprehensible outputs to humans \cite{doi1}, makes CPADF meets the operational requirement ([R1]). Furthermore, CPADF two-level monitoring, accuracy and low false positive/negative rate allows understanding the electricity consumption changes so as to act accordingly ([R2]).

\begin{table}[h!]
\centering
	\caption{Comparison among anomaly detection systems}
\begin{adjustbox}{width=0.99\textwidth}
	\begin{tabular}{|l|c|c|c|c|}
		\hline
		\multicolumn{1}{|c|}{} & Jokar et al. \cite{Jokar2016} & Ford et al. \cite{ford} & Cody et al. \cite{cody} & CPADF \\ \hline
		HD (\%) & 70 & 68.75 & NA & 79.4 \\ \hline
		DR (\%) & 86 & 93.75 & NA & 96 \\ \hline
		FPR (\%) & 16 & 25 & NA & 16.6 \\ \hline
		RMSE & NA & 0.33 & 0.47 & 0.29 \\ \hline
		Anomaly Type & Energy theft & \begin{tabular}[c]{@{}c@{}}Grid overloading\\ Energy theft\end{tabular} & \begin{tabular}[c]{@{}c@{}}Grid overloading\\ Energy theft\end{tabular} & \begin{tabular}[c]{@{}c@{}}Grid overloading\\ Load fluctuation\end{tabular} \\ \hline
	\end{tabular}
	\end{adjustbox}
	\label{compTab}
\end{table}

\begin{figure}
	\begin{center}
		\includegraphics[scale=0.72]{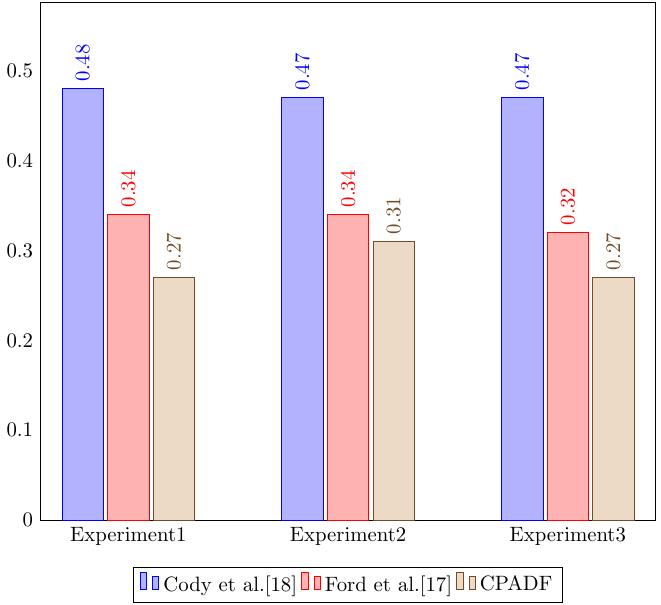}
	\end{center}
	\caption{Comparison of prediction error between CPADF and the state of the art.}
	\label{compa}	
\end{figure}

To the best of our knowledge, there are only three papers \cite{ford,cody,Jokar2016} which have used the same dataset \cite{irish} for AMI anomaly detection. Ford et al. \cite{ford} and Cody et al. \cite{cody} used neural network and decision tree, receptively, to detect two types of energy fraud, whereas Jokar et al \cite{Jokar2016} used SVM based classification to detect energy theft. Since the performance of anomaly detection depends on the accuracy of energy prediction, we first compare the energy prediction performance of CPADF with Ford et al. \cite{ford} and Cody et al. \cite{cody}. For the sake of fairness, we consider the same experiments used in \cite{ford,cody}. The aim of the first experiment is to evaluate to which extent the regression model can predict electricity consumption for the same month a year after the training set, as in \cite{ford,cody}, we exploited August 2009 for training, and August 2010 for validation. Experiment 2 examines the ability to predict electricity consumption the week following several weeks, as in \cite{ford,cody} we considered weeks from September. To evaluate the ability of electricity prediction within the same weather season, experiment 3 uses electricity consumption from June 2010 for training, then validated the model on July of the same year. The three experiment results are presented in Figure \ref{compa}, as we can notice CPADF provides the lowest root mean squared error for the three experiments. 

Table \ref{compTab} displays a comparison between the anomaly detection overall performances of CPADF, \cite{ford}, \cite{cody} and \cite{Jokar2016}. As we can see, CPADF provides the best detection rate and the lowest prediction error (RMSE), however CPADF presents 0.6\%  of extra FPR in comparison with \cite{Jokar2016}. The proposed system in \cite{Jokar2016} uses synthetic malicious samples to build the system, which may cause FNR to increase when the malicious pattern changes, because the classifier would not detect attack types that deviate significantly from the synthetic malicious samples used to train the system.

\section{Conclusion}\label{sec:conc}
In this paper, grid overloading cyberattacks in the context of smart grid AMI are considered. These cyberattacks aim at increasing the energy usage and load fluctuation to disturb the power grid and cause a large area blackout. After analyzing them, CPADF a distributed anomaly detection system based on regression decision trees is proposed. CPADF relies on the predictability of smart home and neighborhood consumption patterns. We showed that CPADF can detect grid overloading cyberattacks regardless of the strategy employed by the attacker and with an optimal detection delay. The simulation results on a real dataset of 500 customers demonstrate that CPADF provides a high detection rate and a low false positive rate with short running time and memory requirement. As future work, we need to explore more cyberattacks and to improve the anomaly detection algorithm using more sophisticated machine learning methods.

\clearpage 

\bibliographystyle{elsarticle-num}
\bibliography{mybibfile}

\end{document}